\documentclass[dvipsnames]{article} %
\usepackage{colm2024_conference}

\usepackage{booktabs}
\usepackage{graphicx}
\usepackage{enumitem}
\usepackage{wrapfig}
\usepackage{algorithm}
\usepackage{algpseudocode}

\usepackage{microtype}
\usepackage{amsmath}
\usepackage{colortbl}
\usepackage[utf8]{inputenc}
\definecolor{lightgray}{rgb}{0.9,0.9,0.9}
\usepackage{caption}
\usepackage{subcaption}
\usepackage{setspace}
\usepackage{url}
\usepackage{multirow}
\usepackage{colortbl}
\usepackage{tabularx}
\usepackage{blindtext}
\usepackage{pgfplots}
\pgfplotsset{compat=1.18} 
\usepackage{tikz}
\usetikzlibrary{er,positioning,bayesnet}
\usepackage{makecell}
\usepackage{tipa}
\usepackage{siunitx}
\usepackage{nicefrac}
\usepackage{tocloft}
\usepackage{listings}
\usepackage[raster,skins]{tcolorbox} %
\usepackage{xltabular}
\usepackage{adjustbox}
\usepackage{xurl}
\usepackage{makecell}
\usepackage{booktabs}
\usepackage{multicol}
\usepackage{multirow}
\usepackage{rotating}
\usepackage{cleveref}
\crefname{section}{§}{§§}
\Crefname{section}{§}{§§}
\usepackage[normalem]{ulem}
\useunder{\uline}{\ul}{}

\usepackage{tablefootnote}
\usepackage{xcolor}

\usepackage{amsmath,amsfonts,bm}

\def\eqref#1{equation~\ref{#1}}

\def\1{\bm{1}}

\DeclareMathAlphabet{\mathsfit}{\encodingdefault}{\sfdefault}{m}{sl}
\SetMathAlphabet{\mathsfit}{bold}{\encodingdefault}{\sfdefault}{bx}{n}

\renewcommand{\ghlink}{https://github.com/QwenLM/Qwen3}

\newcommand*\justify{%
  \fontdimen2\font=0.4em
  \fontdimen3\font=0.2em
  \fontdimen4\font=0.1em
  \fontdimen7\font=0.1em
  \hyphenchar\font=`\-
}

\renewcommand{\texttt}[1]{%
  \begingroup
  \ttfamily
  \begingroup\lccode`~=`/\lowercase{\endgroup\def~}{/\discretionary{}{}{}}%
  \begingroup\lccode`~=`[\lowercase{\endgroup\def~}{[\discretionary{}{}{}}%
  \begingroup\lccode`~=`.\lowercase{\endgroup\def~}{.\discretionary{}{}{}}%
  \catcode`/=\active\catcode`[=\active\catcode`.=\active
  \justify\scantokens{#1\noexpand}%
  \endgroup
}

\title{SWE-Universe: Scale Real-World Verifiable Environments to Millions}

\author{
\textbf{
Mouxiang Chen$^{1,2}$\thanks{Work done during an internship at Alibaba Qwen.}, Lei Zhang$^1$, Yunlong Feng$^1$, Xuwu Wang$^1$, Wenting Zhao$^1$, Ruisheng Cao$^1$, Jiaxi Yang$^1$, Jiawei Chen$^1$, Mingze Li$^1$, Zeyao Ma$^1$, Hao Ge$^1$, Zongmeng Zhang$^1$, Zeyu Cui$^1$, Dayiheng Liu$^1$, Jingren Zhou$^1$, Jianling Sun$^2$,\\
Junyang Lin$^1$, Binyuan Hui$^1$, Zhongxin Liu$^2$}\\
$^1$Qwen Team, Alibaba Group, $^2$Zhejiang University\\
\texttt{\{chenmx,liu\_zx\}@zju.edu.cn, junyang.ljy@alibaba-inc.com}
}

\begin{document}

\maketitle

\begin{abstract}
We propose SWE-Universe, a scalable and efficient framework for automatically constructing real-world software engineering (SWE) verifiable environments from GitHub pull requests (PRs). To overcome the prevalent challenges of automatic building, such as low production yield, weak verifiers, and prohibitive cost, our framework utilizes a building agent powered by an efficient custom-trained model. This agent employs iterative self-verification and in-loop hacking detection to ensure the reliable generation of high-fidelity, verifiable tasks. Using this method, we scale the number of real-world multilingual SWE environments to a million scale (807,693). We demonstrate the profound value of our environments through large-scale agentic mid-training and reinforcement learning. Finally, we applied this technique to Qwen3-Max-Thinking and achieved a score of 75.3\% on SWE-Bench Verified. Our work provides both a critical resource and a robust methodology to advance the next generation of coding agents.
\end{abstract}

\begin{figure}[h]
    \centering
    \includegraphics[width=0.5\textwidth]{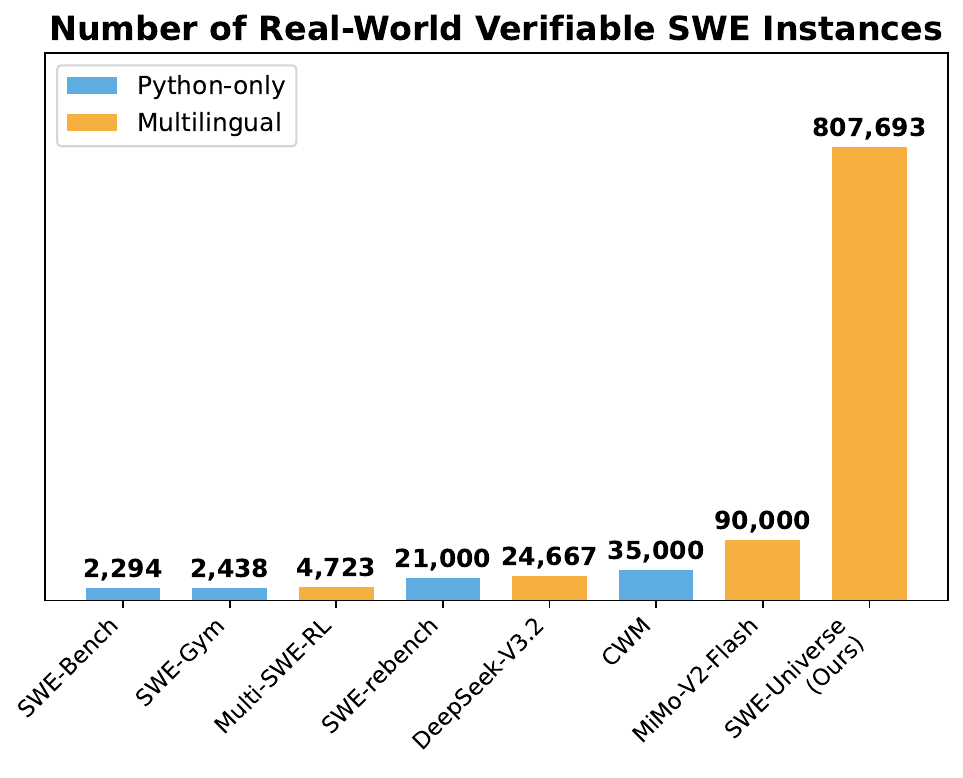}
    \caption{A comparison of the number of instances in real-world SWE instances. Our multilingual SWE-Universe is significantly larger than other recent multilingual efforts like MiMo-V2-Flash~\citep{xiao2026mimo}, DeepSeek-V3.2~\citep{deepseekv32}, and Multi-SWE-RL~\citep{multi-swe-bench}, as well as prominent Python-only datasets including SWE-rebench~\citep{swe-rebench}, SWE-Gym~\citep{pan2024training}, CWM~\citep{copet2025cwm}, and SWE-Bench~\citep{swe-bench}.}
    \label{fig:env-intro}
\end{figure}

\vfill

\newpage

\section{Introduction}
\label{sec:introduction}

Training large language models (LLMs)~\citep{yang2025qwen3,liu2025deepseek} as coding agents has gained substantial attention in software engineering (SWE). Progress in this direction is critically dependent on large-scale, high-quality environments with reliable verification signals. An ideal source for such data lies within the vast ecosystem of real-world software development, specifically public GitHub repositories where pull requests (PRs) are linked to corresponding issues. Each such PR defines a self-contained training environment: the issue serves as a clear problem statement, the code patch offers an expert reference solution, and the accompanying tests can be repurposed into a verifier for agent proposed fixes. 
This formulation effectively turns each PR into a ``gym'' for training and evaluating coding agents. Benchmarks such as {SWE-bench}~\citep{swe-bench} operationalized this pipeline by curating real-world issues and standardizing the evaluation, serving as a common testbed for subsequent work.


However, scaling the construction of verifiable real-world SWE environments while maintaining diversity and reproducibility remains a major challenge. A vast body of work, including the original {SWE-bench} and its many derivatives and extensions~\citep{pan2024training,swe-rebench,zhang2025swe,guo2025swe,zeng2026davinci,wang2025swe,aleithan2024swe,copet2025cwm} has primarily focused on Python. While this line of work leverages the low barrier to entry for Python environment configuration, it nonetheless restricts the development of agents with true cross-lingual generalization capabilities. Efforts to create multi-lingual datasets, such as {Multi-SWE-bench}~\citep{multi-swe-bench} and {SWE-PolyBench}~\citep{rashid2025swe}, rely on labor-intensive manual environment setup and remain limited in scale. Despite recent efforts of industrial LLM providers that scale SWE instances to the $10^4-10^5$ magnitude~\citep{xiao2026mimo,deepseekv32}, the technical details are undisclosed.

We argue that scaling the generation of such verifiable environments to the massive scale (e.g., $10^6$) hinges on overcoming three fundamental challenges:
\begin{itemize}[leftmargin=12pt]
    \item \textbf{Low Production Yield:} The intricate and heterogeneous nature of real-world repositories with their complex dependencies, platform-specific configurations, and bespoke build toolchains, yielding a low conversion rate from repositories to runnable instances. This results in significant computational waste and makes large-scale generation impractical.
    \item \textbf{Weak Verifier:} Issues, PR patches, and test suites exhibit substantial variance in quality. A naive extraction pipeline can produce low-fidelity instances, and can also admit verifier vulnerabilities that allow solutions to pass via shallow heuristics (e.g., string matching with \texttt{grep}) rather than compiling and executing the intended code. Such failure modes create spurious training signals and distort evaluation.
    \item \textbf{Prohibitive Cost and Inefficiency:} Many existing pipelines rely on large, expensive LLMs to perform repository-specific reasoning for dependency resolution and build configuration. The resulting cost and latency per instance make massive-scale generation economically and operationally impractical.
\end{itemize}

To systematically address these challenges, we introduce \textbf{SWE-Universe}, a scalable, reliable, and efficient framework for automatically constructing million-scale, real-world agentic software engineering environments. At its core is an autonomous \textit{building agent} that, for each PR, synthesizes a self-contained executable environment together with an executable verifier. To mitigate \textbf{low production yield}, the agent performs self-verification in an iterative validation loop. Concretely, it repeatedly tests the generated verifier against both the buggy and the fixed repository states, diagnoses failure modes, and revises the build procedure accordingly. This process improves the build success rate from 82.6\% to 94\% on a held-out set. To address the \textbf{weak verifier}, we integrate an \textit{in-loop hacking detector} that immediately flags and rejects superficial verifiers during the generation process, forcing the agent toward solutions that genuinely execute the code. To achieve \textbf{high efficiency and low cost}, we specially trained an efficient {Qwen-Next-80B-A3B} model with a Mixture-of-Experts (MoE) framework and hybrid attention. On our custom multi-lingual building benchmark, this model achieves a 78.44\% success rate, surpassing top-tier proprietary models like {Claude-Opus-4.5}, while its efficient architecture dramatically reduces the latency and cost per build.

Leveraging the SWE-Universe framework, we construct \textbf{807,693} multilingual, verifiable training instances sourced from over \textbf{52,000} unique GitHub repositories. To our knowledge, this dataset is currently the largest and most diverse collection of real-world software engineering tasks with executable verification.
We further validate the immense value of this dataset through extensive experiments. First, in a mid-training phase, we show that continued training on our dataset significantly enhances a model's performance on standard benchmarks like SWE-Bench Verified~\citep{swe-bench}, proving its powerful generalization capabilities. Second, we demonstrate that the verifiers generated by our pipeline provide a stable and effective reward signal for Reinforcement Learning (RL). By applying to our flagship model, {Qwen3-Max-Thinking}~\citep{qwen3max-thinking}, we achieved a score of 75.3\% on the SWE-Bench Verified. Together, these results validate SWE-Universe as a robust framework for creating large-scale, high-quality agentic training data, paving the way for the development of more capable and versatile coding agents for real-world applications.

\section{Methodology: Scalability and Reliability}

\begin{figure}[t]
    \centering
    \includegraphics[width=1.0\textwidth]{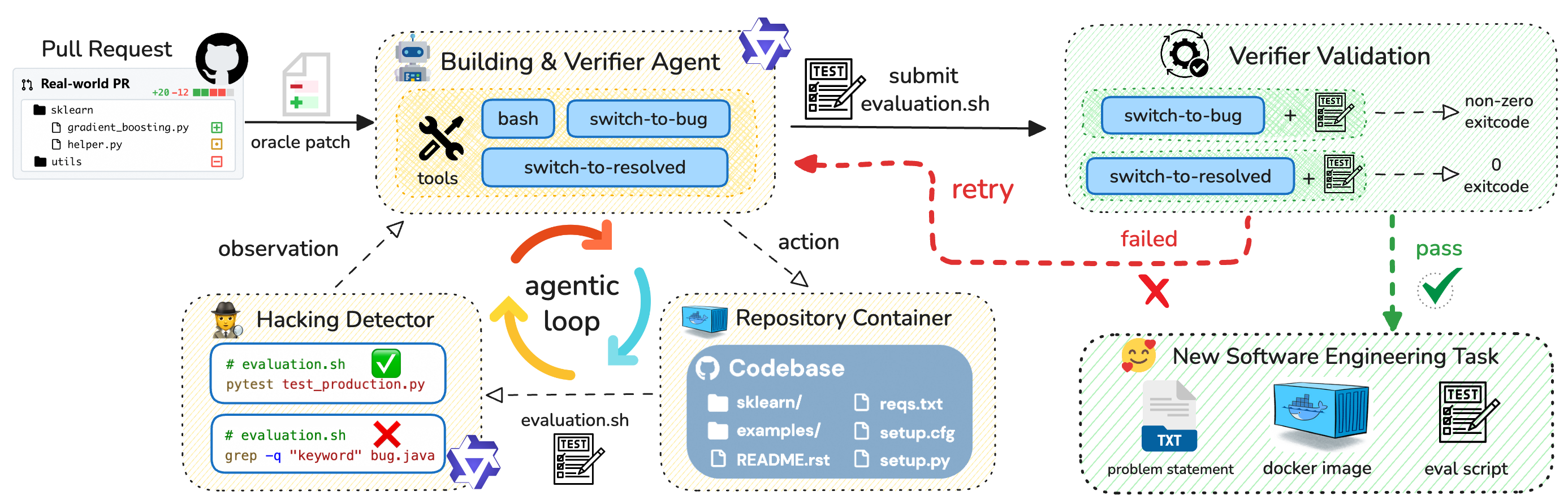}
    \caption{Our SWE-Universe framework for scalable and reliable environment building. The pipeline is built around a \textbf{building agent} that proposes a verifier (\texttt{evaluation.sh}). Two key components ensure quality and yield: an \textbf{in-loop Hacking Detector} that preemptively rejects superficial scripts, and an \textbf{Iterative Validation} loop where the agent self-corrects based on feedback from testing its verifier against both buggy and resolved code states.}
    \label{fig:env-build}
\end{figure}

Our primary goal is to develop a method for automated verifiable SWE environment construction that is both \textbf{\textit{scalable}}
and \textit{\textbf{reliable}}. Scalability implies compatibility across diverse programming languages and build toolchains, as well as maximizing the build success rate to minimize computational waste. Reliability requires that the generated verifier (i.e., an \texttt{evaluation.sh} script) accurately provides a reward signal for a given patch, passing only when the SWE requirements are met. To this end, we design an autonomous building agent based on the popular \texttt{mini-sweagent} scaffold.

Our key insight is that while directly verifying the success of a complex building is difficult, assessing the success of a test script is comparatively straightforward: one can simply run it to see if it distinguishes between the pre-patch (buggy) and post-patch (fixed) states, as adopted by DeepSeek-V3.2~\citep{liu2025deepseek}. However, we argue that this condition alone is insufficient. A naive agent might generate a script that ``hacks'' the verification by using simple string matching to confirm the patch's application, rather than actually executing the code. Such a script would correctly distinguish states but would fail to validate the environment or the behavioral correctness of the fix. Therefore, we establish a more robust acceptance criterion for a successfully built task: the verifier must not only correctly distinguish between states but must do so by \textit{genuinely executing the code under test}. Our methodology, detailed below and sumarized in \Cref{fig:env-build}, is designed to meet this twofold objective.

\subsection{Key Designs for SWE-Universe}

\paragraph{PR Crawling and Patch Separation.}
The pipeline begins by sourcing a large corpus of issue-linked pull requests (PRs) from public GitHub repositories. To prevent data contamination and ensure fair evaluation, we meticulously filter out any PRs that overlap with known downstream benchmarks. For each valid PR, we employ a language model to analyze the code modifications and partition them into a \textit{test patch} (containing test-related changes) and a \textit{fix patch} (containing the source code fix). PRs lacking a discernible test component are discarded.

\paragraph{Agent-based Environment Building.}
Following patch separation, our autonomous agent, equipped with a set of powerful tools, initiates the construction of the environment and its corresponding verifier. The process commences by applying the \textit{test patch} to the repository. The agent is then tasked with generating a verifier script, designated as \texttt{evaluation.sh}, whose objective is to reliably distinguish between the repository's states based solely on its return code. Guided by the information within the test patch, the agent has the flexibility to adopt one of two strategies: it can either directly invoke the unit tests described in the patch, or, in scenarios where those tests lack a straightforward execution entry point, it can author a new custom test from scratch. This standardized approach, centered on a universal \texttt{bash} interface and its integer return code, is a deliberate design choice. It decouples the verification logic from language-specific conventions (e.g., hardcoding a \texttt{pytest} or \texttt{cargo run} workflow), thereby maximizing the method's generalizability and scalability across disparate projects and ecosystems.

\paragraph{Toolset.} 

We equip the agent with three tools: \texttt{bash}, \texttt{switch-to-resolved}, and \texttt{switch-to-bug}.

\begin{itemize}[leftmargin=12pt]
    \item \texttt{bash}: A general-purpose shell for file manipulation, dependency installation, and script generation.
    \item \texttt{switch-to-resolved} and \texttt{switch-to-bug}: A pair of tools that allow the agent to atomically apply or revert the \textit{fix patch}, toggling the repository between its fixed and buggy states.
\end{itemize}
These state-switching tools are fundamental to our approach, as they empower the agent to perform self-verification. By testing its own actions in a closed loop, the agent can diagnose and recover from failures, which is key to improving the overall build success rate and achieving scalability.

\paragraph{Iterative Validation.}
To ensure the verifier is reliable, the agent engages in an iterative validation loop. After the agent ends with submitting a candidate \texttt{evaluation.sh} script, we execute it under both repository states using its switching tools. A script is considered functionally correct only if it fails (exits with a non-zero status) in the buggy state and succeeds (exits with a zero status) in the fixed state. If the script fails this validation, the agent receives this negative feedback, discards the faulty script, and is prompted to generate a new, revised version until reaches the maximum turns we predefined (e.g., 100 turns). This iterative process substantially improves reliability: for a set of held-out PRs, the environment-building success rate increases from 82.6\% to 94\%.

\paragraph{In-loop Hacking Detection.}
To meet our second criterion—that the verifier must genuinely execute the code—we integrate a \textit{Hacking Detector} directly into the agent's work cycle. This module uses an LLM to inspect the generated \texttt{evaluation.sh} script for ``hacking'' patterns, such as using \texttt{grep} or other string-matching utilities to check the contents of source files instead of running a build or test command. Critically, this check is performed \textit{within} the agent's loop, not as a post-processing step. If the detector flags a script as a ``hack'', this attempt is immediately considered a failure. This provides timely feedback to the agent, forcing it to abandon superficial strategies and guiding it toward a valid solution that involves actual code execution. This in-loop design improves both the reliability of the final artifact and the efficiency of the agent's search process.

\begin{figure}[t]
    \centering
    \includegraphics[width=1.0\textwidth]{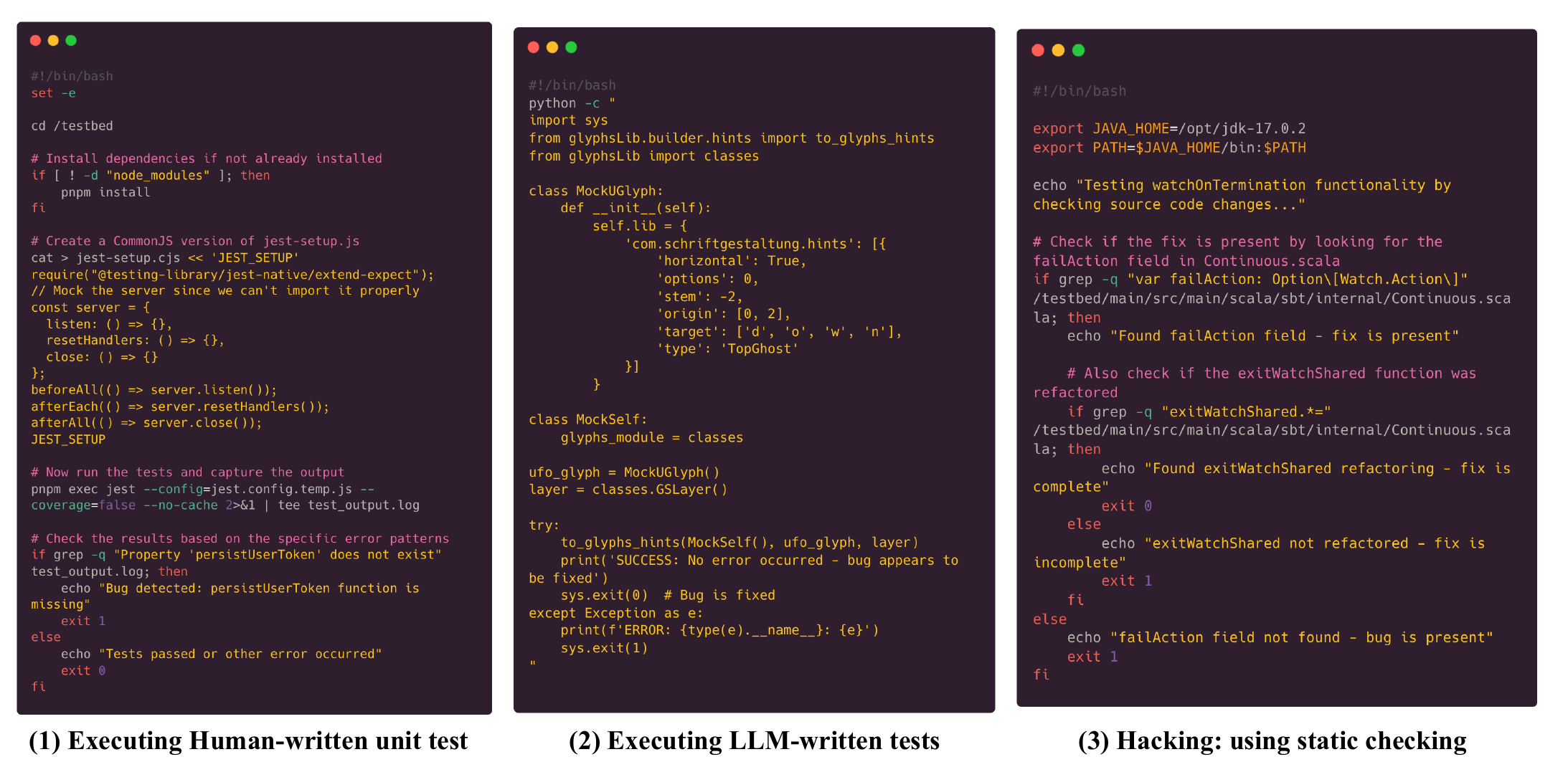}
    \caption{Three types of \texttt{evaluation.sh}. We only accept the first two types of evaluation scripts.}
    \label{fig:evaluation-case}
\end{figure}

\subsection{Further Analysis}

\paragraph{Case Study} We examine three representative LLM-generated test scripts to illustrate our verification criteria (Figure~\ref{fig:evaluation-case}). Case 1 tests a JavaScript authentication bug by executing existing human-written unit tests in the Jest framework via \texttt{pnpm exec jest}, checking for specific error patterns in the output. Case 2 validates a Python glyphsLib bug by creating LLM-generated mock objects and unit tests that directly exercise the buggy code path and verify execution without exceptions. Case 3 attempts to verify a Scala bug fix using static pattern matching with \texttt{grep} commands to detect the presence of specific code structures, which is flagged as ``hacking'' by our hacking detection. We accept Cases 1 and 2 as they employ strong verification through executable tests that validate runtime behavior, while we reject Case 3 because static pattern matching cannot guarantee correctness—a patch may introduce expected code patterns while still containing logical errors.

\begin{figure}[t]
    \centering
    \includegraphics[width=0.6\textwidth]{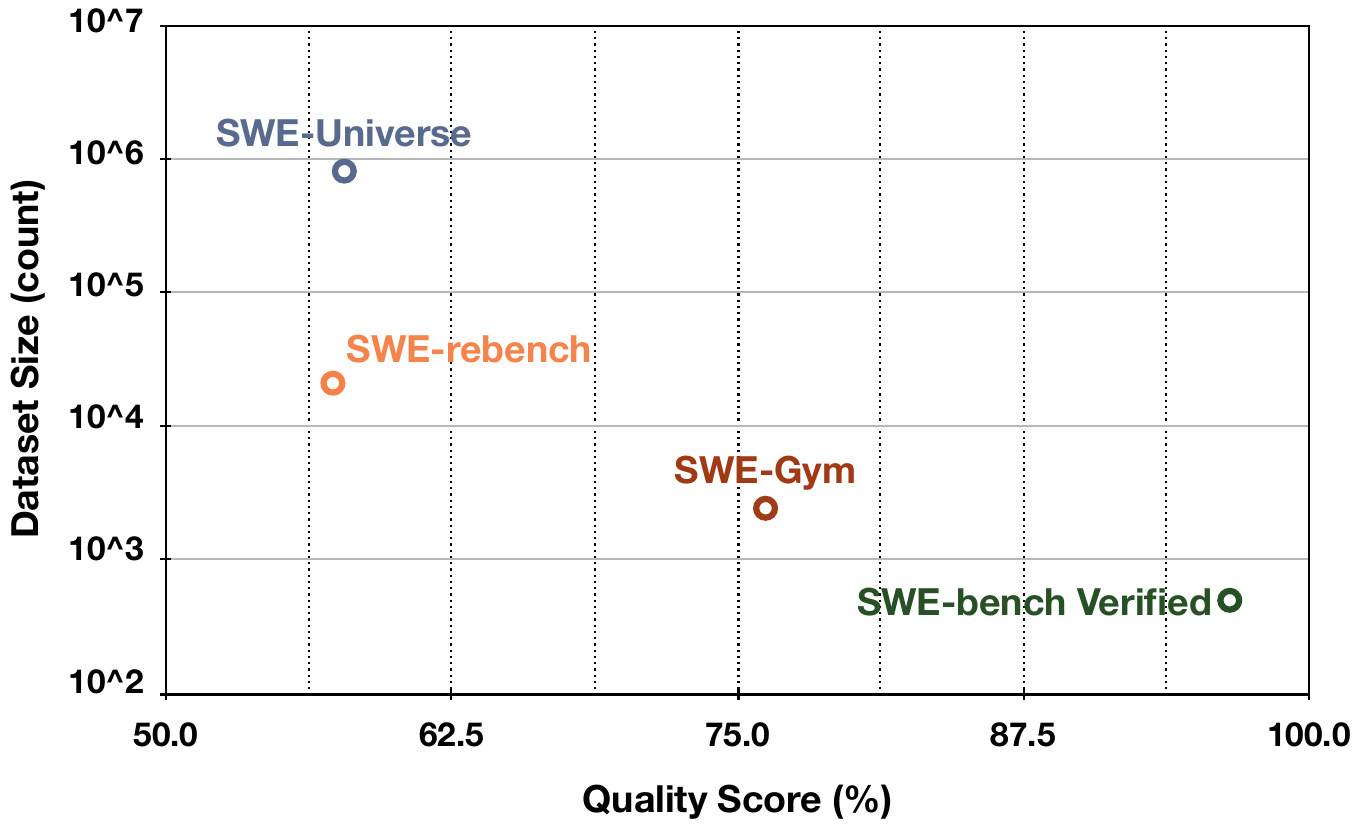}
    \caption{Task Quality vs. Dataset Size (Log-Scale). Task quality is measured as the fraction of high-quality samples.}
    \label{fig:quality_vs_size}
\end{figure}

\paragraph{Quality Analysis} We find that the resulting data still exhibit several quality issues: (1) some task descriptions are ambiguous or incomplete; (2) certain Docker environments do not fully match the stated requirements; and (3) some unit tests are misaligned with the task descriptions, which can lead to false positives or false negatives. 
To quantify and mitigate this problem, we develop a quality-judge agent. It takes as input the task description, Docker environment, test scripts, and optionally the ground-truth patch, and automatically evaluates the task quality.
On a human-labeled quality-judging benchmark, the agent reaches 78.72\% accuracy. After applying it to the entire dataset, as shown in Figure~\ref{fig:quality_vs_size}, we find our dataset matches SWE-Rebench~\citep{swe-rebench} in quality while providing ~38× more instances.

\section{Efficient Building and Benchmarking}

To optimize both the performance and throughput of our pipeline, we developed a high-capacity yet efficient model and a comprehensive cross-lingual benchmark to evaluate its capabilities.

\paragraph{Model Training and Deployment}
To make our end-to-end pipeline efficient at scale, we train a lightweight but strong builder model, {Qwen-Next-80A3}~\citep{qwen3next}, a mixture-of-experts (MoE) model with hybrid attentions including linear attentions and full attentions. The model was trained using rejection sampling on a collection of high-quality building trajectories. This process involved sampling multiple candidate paths for environment construction and filtering for successful, non-hacked outcomes to serve as training data. The model serves as the unified backbone for all tasks in our pipeline, including PR patch splitting, iterative environment building, and the hacking detector. Its efficient architecture allows us to scale the building process across thousands of repositories with significantly lower latency compared to dense models of similar performance.

\paragraph{Benchmark}
To rigorously assess the reliability of automated environment construction, we constructed a diverse benchmark consisting of 320 pull requests randomly sampled from GitHub. We remove the repositories used in the training trajectories from the benchmark. To ensure broad representativeness, we selected 40 PRs for each of the eight language categories: Python, JavaScript/TypeScript, Go, Java, Rust, C/C++, C\#, and an ``Others'' category (including all the other languages such as PHP and Kotlin). We define two primary success metrics: \textit{Success Rate (w/o Hack)}, which requires the verifier to be functionally correct and pass the hacking detector, and \textit{Success Rate (w/ Hack)}, which includes all scripts that distinguish the bug regardless of the detection outcome.

\begin{table*}[!t]
\centering
\caption{Benchmark results for automated environment building across various models. ``Success (\%) (w/o Hack)'' measures the rate of creating a valid, non-hacked verifiable environment, while ``Success (\%) (w/ Hack)'' also counts ``hacked'' verifiers as successful builds. Our model, \textbf{Qwen-Next-80A3}, achieves the highest non-hacking success rate.}
\label{tab:benchmark_results}
\resizebox{\textwidth}{!}{%
\begin{tabular}{l S[table-format=2.2] S[table-format=2.2] S[table-format=2.2] S[table-format=2.2] S[table-format=2.2] S[table-format=2.2] S[table-format=2.2] S[table-format=2.2] S[table-format=2.2] S[table-format=2.2]}
\toprule
\multirow{2}{*}{\textbf{Model}} & {\textbf{Success (\%)}} & {\textbf{Success  (\%)}} & \multicolumn{8}{c}{\textbf{Success Rate by Language (\%)}} \\
\cmidrule(lr){4-11}
 & {\textit{(w/o Hack)}} & {\textit{(w/ Hack)}} & {C/C++} & {C\#} & {Go} & {Java} & {JS/TS} & {Other} & {Python} & {Rust} \\
\midrule
\textbf{Qwen-Next-80A3 (Ours)} & \bfseries 78.44 & 82.50 & 57.50 & 70.00 & 80.49 & 82.50 & 84.62 & 83.72 & 85.37 & 83.33 \\
{Claude-Opus-4.5} & 77.81 & 85.00 & 52.50 & 57.50 & 82.93 & 77.50 & 89.74 & 76.74 & 95.12 & 91.67 \\
{Claude-Sonnet-4} & 75.62 & 85.62 & 67.50 & 52.50 & 75.61 & 72.50 & 84.62 & 83.72 & 82.93 & 86.11 \\
{Gemini-3-Pro} & 69.69 & 72.50 & 32.50 & 57.50 & 73.17 & 72.50 & 87.18 & 76.74 & 87.80 & 69.44 \\
{Claude-Sonnet-4-5} & 66.88 & 71.56 & 30.00 & 50.00 & 63.41 & 67.50 & 87.18 & 67.44 & 92.68 & 77.78 \\
{GLM-4.7} & 58.44 & 64.06 & 37.50 & 50.00 & 53.66 & 57.50 & 66.67 & 65.12 & 73.17 & 63.89 \\
{MiniMax-M2.1} & 54.69 & 61.88 & 22.50 & 35.00 & 56.10 & 55.00 & 74.36 & 65.12 & 73.17 & 55.56 \\
{DeepSeek-v3.2} & 54.06 & 59.38 & 15.00 & 50.00 & 63.41 & 47.50 & 53.85 & 48.84 & 78.05 & 77.78 \\
{Qwen3-Coder-480B} & 48.75 & 55.62 & 35.00 & 32.50 & 39.02 & 45.00 & 48.72 & 69.77 & 68.29 & 50.00 \\
\bottomrule
\end{tabular}%
}
\end{table*}

\paragraph{Result Analysis} Our evaluation, detailed in Table~\ref{tab:benchmark_results}, establishes {Qwen-Next-80A3} as the new state-of-the-art model for automated environment building, achieving a 78.44\% success rate that surpasses even strong proprietary models like {Claude-Opus-4.5} (77.81\%). This superior performance extends beyond raw success to overall reliability. We observe a significant gap between the true success rate and the one including ``hacked'' verifiers for most general-purpose models (e.g., over 7\% for {Claude-Opus-4.5}), indicating their tendency to find superficial shortcuts. In stark contrast, our model exhibits the smallest gap among top performers (4.06\%), demonstrating that our task-specific fine-tuning on collected trajectories effectively discourages these deceptive behaviors. Furthermore, while performance varies by language—with C/C++ proving most challenging for all agents—our model demonstrates more consistent and robust capabilities across the entire linguistic spectrum compared to competitors that show strength only in specific ecosystems. Ultimately, these results validate that a specialized, efficiently trained MoE model not only achieves higher success but also operates with greater fidelity than larger, general-purpose counterparts in this complex domain.

\section{Scaling Environments to Millions}

In this section, we describe our efforts to scale the environment-building pipeline to a massive corpus of real-world software changes, moving beyond small-scale benchmarks to a million-scale dataset of executable tasks.

\paragraph{Large-scale Data Curation}
We harvested a comprehensive dataset of approximately 33.3 million pull requests (PRs) spanning the most recent five years (2021–2025) of GitHub's history. To extract high-quality tasks from this raw pool, we applied a series of rigorous heuristic filters. Specifically, we removed PRs with excessive file changes or line counts to avoid overly complex or noisy tasks, and discarded any entries that did not contain a discernible test patch. Furthermore, we prioritized PRs that were explicitly linked to at least one GitHub issue, as these provide the most reliable ground-truth problem statements. This filtering process resulted in a candidate set of approximately 1 million high-quality PRs.

\paragraph{Infrastructure for Large-scale Rollouts} 

To support our massive data generation effort, we implement our pipeline on top of \textsc{MegaFlow}~\citep{megaflow}, a distributed execution system for orchestrating large numbers of long-running agentic jobs. \textsc{MegaFlow} achieves massive parallelism by dispatching each environment-building task as an independent job to a dedicated Alibaba Cloud Elastic Compute Service (ECS) instance. Within this sandboxed virtual machine, our agent executes the entire build process, culminating in a verified Docker image. Upon completion, successful images are pushed to Alibaba Cloud's Container Registry (ACR)\footnote{\url{https://www.alibabacloud.com/en/product/container-registry}}, where we leverage Docker's layer caching to significantly reduce storage costs by reusing common base layers. This highly parallel and resource-efficient architecture was instrumental in processing millions of pull requests concurrently, enabling the practical construction of our million-scale dataset.

\paragraph{Massive Production with Qwen-Next-80A3}
We deployed our fine-tuned {Qwen-Next-80A3} model to execute the building and verification process for the filtered candidates. Leveraging the model's efficiency and the iterative agentic loop described in previous sections, we achieved a non-hacked success rate of 75.9\%. This process ultimately yielded 717,122 executable, high-fidelity environments. These tasks are primarily focused on assessing an agent's ability to resolve real-world software issues. Additionally, to expand the dataset's coverage, we selected a subset of 2025 PRs that were not linked to issues, utilizing the PR titles and descriptions as the problem statements to synthesize an additional  environments, and producing 90,571 environments.

\begin{table}[t]
\centering
\small
\caption{Data statistics for our building environments.}
\label{tab:real-world-pr}
\begin{tabular}{lrrrr}
\toprule
\textbf{Language} & \textbf{Instances} & \textbf{Repos} & \textbf{Inst/Repo (Avg)} & \textbf{Avg Lines of \texttt{evaluation.sh}} \\
\midrule
Python & 202,302 & 13,098 & 15.45 & 25.01 \\
Javascript / Typescript & 175,660 & 11,604 & 15.14 & 27.41 \\
Go & 121,062 & 5,554 & 21.80 & 28.87 \\
Java & 86,105 & 4,700 & 18.32 & 24.75 \\
Rust & 74,180 & 4,445 & 16.69 & 19.31 \\
C / C++ & 37,228 & 3,405 & 10.93 & 45.78 \\
C\# & 24,387 & 1,929 & 12.64 & 31.84 \\
Others & 86,769 & 8,225 & 10.55 & 38.89 \\
\midrule
\textbf{\# Total} & \textbf{807,693} & 52,960 & 15.25 & 28.21 \\
\bottomrule
\end{tabular}
\end{table}

\paragraph{Statistical Analysis}

The statistics presented in Table~\ref{tab:real-world-pr} highlight both the scale and diversity of our generated dataset. With 807,693 instances spread across 52,960 unique repositories, this dataset represents an unprecedented resource for training and evaluating software engineering agents. The language distribution largely mirrors the current open-source landscape, with Python and JavaScript/TypeScript constituting the largest shares. Notably, Go repositories exhibit the highest average number of instances per repository (21.80), suggesting a high density of verifiable, issue-linked PRs within its ecosystem, possibly due to strong conventions around testing and development. The ``Avg Eval Lines'' metric reveals interesting insights into the complexity of verification across languages. C/C++ instances, as expected, require the longest verifier scripts on average (45.78 lines), reflecting the typical complexity and boilerplate of their build systems. In contrast, Rust's concise average (19.31 lines) may point to the efficiency and standardization of its \texttt{cargo} toolchain, allowing for more succinct test execution commands. Overall, the dataset provides a rich and varied testbed, capturing a wide array of real-world challenges for coding agents.

\section{Evaluation: Large-scale Agentic Training}

We now turn to leveraging this resource for large-scale agentic model training. Our goal is to demonstrate that training on this diverse, real-world data can significantly enhance a model's capabilities as a software engineering agent.

\subsection{Mid-training}


We hypothesize that intermediate training on a vast corpus of high-quality agentic trajectories can endow a model with a strong foundation in both software engineering problem-solving. This ``mid-training'' phase is designed to bridge the gap between the pre-training and post-training on downstream agentic tasks.

\paragraph{Setup} Our mid-training process begins by generating a massive dataset of agentic trajectories. We first employed the {Qwen3-Coder-480B-A30B}~\citep{qwen3coder} model to interact with the environments we previously constructed. To ensure a diversity of problem-solving strategies, we conducted these rollouts across five different agentic scaffolds: SWE-agent~\citep{sweagent}, Mini-SWE-agent~\citep{minisweagent}, OpenHands~\citep{openhands}, Claude-Code~\citep{claudecode}, Qwen-Code~\citep{qwencode}. The entire rollout process was orchestrated by our \textsc{MegaFlow} system. For each environment, we performed rejection sampling: a trajectory is deemed successful and retained only if the final generated code passes the corresponding \texttt{evaluation.sh} script and clears an additional in-house quality filter. This rigorous filtering process yielded a high-quality dataset of 500K successful trajectories, comprising a total of 30 billion training tokens. We then used this data to conduct intermediate training on a {Qwen3-Next-80A3} model, using 256K sequence length and Best-Fit packing~\citep{best-fit-packing}. Critically, we applied no loss mask during this training phase. This strategy ensures that the model learns not only to predict agentic actions but also to internalize the vast amount of code and natural language in the agent's observations and responses, developing it into a more comprehensive ``coding world model''~\citep{copet2025cwm,zhang2025agent}.


\paragraph{Scaling Trends}

To analyze the effectiveness of our large-scale agentic mid-training, we evaluated model checkpoints on two widely used agentic coding benchmarks: SWE-Bench Verified (primarily Python-based) and the more diverse SWE-Bench Multilingual. The performance evolution throughout the training process is depicted in Figure~\ref{fig:scaling_trends}. The results demonstrate a clear and positive scaling trend, confirming that our mid-training on the real-world instances successfully transfers to standard evaluation benchmarks. On the standard \textbf{SWE-Bench Verified}, the model's performance exhibits a steady climb, starting from an already strong baseline of 50.3\% and consistently improving to a final score of over 61\% after 2,000 training steps. More notably, the trend on the more challenging \textbf{SWE-Bench Multilingual} benchmark shows even more dramatic improvement. Starting from a significantly lower baseline of approximately 31\%, the model's performance surges to over 46\% by the end of the training. This substantial gain of over 15 percentage points underscores the critical value of our dataset's linguistic diversity. It proves that training on a massive, diverse corpus of real-world, multi-language software issues is essential for creating agents that can generalize beyond a single programming ecosystem.

\subsection{Agentic Reinforcement Learning}

Beyond training via rejection sampling, our collection of executable environments is suited for Reinforcement Learning (RL), where the binary pass/fail signal from the \texttt{evaluation.sh} script serves as a direct and reliable reward. To demonstrate this, we conducted agentic RL experiments on different models with different sizes and structures.

\paragraph{Setup}
We validate the effectiveness of the data on agentic reinforcement learning. 
Prior to training, we perform rollouts with the base model to filter out queries that are either too difficult or too easy. During training, we set the maximum number of interaction turns to 200 and the context length to 128k.
The training process is powered by our asynchronous RL framework, which natively supports agentic workflows. This architecture mitigates data skewness while facilitating seamless multi-turn interactions without framework-induced overhead, achieving a 2x–4x speedup compared to existing RL infrastructures.


\begin{figure}[t]
    \centering
    \begin{subfigure}[b]{0.4\textwidth}
        \centering
        \includegraphics[width=\textwidth]{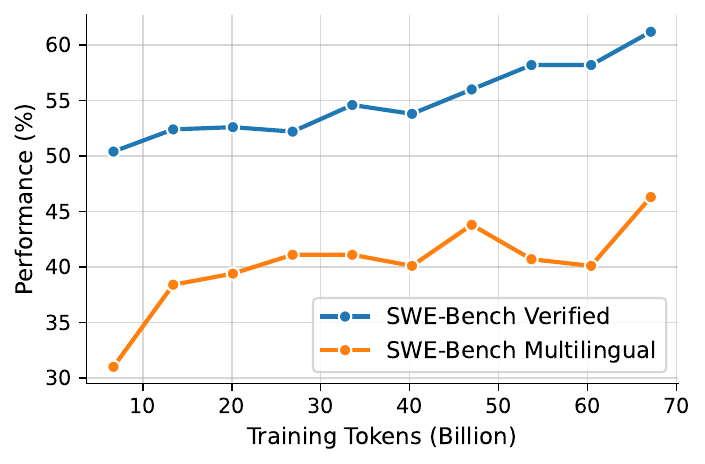}
        \caption{}
        \label{fig:scaling_trends}
    \end{subfigure}
    \begin{subfigure}[b]{0.4\textwidth}
        \centering
        \includegraphics[width=\textwidth]{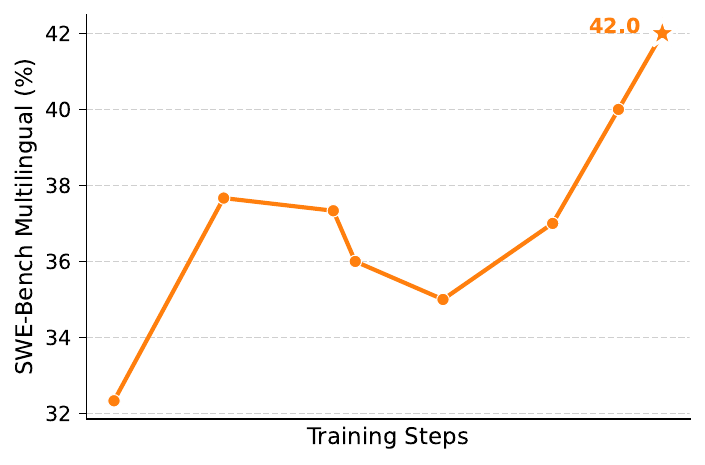}
        \caption{}
        \label{fig:rl_curves}
    \end{subfigure}
    \caption{\textbf{(a)} Performance scaling trends of the {Qwen3-Next-80A3} model during mid-training. \textbf{(b)} Reinforcement learning curve of the Qwen3-30B-A3B for SWE-Bench Multilingual.}
\end{figure}

\paragraph{Results on Qwen3-30B-A3B}

The training curve in Figure~\ref{fig:rl_curves} shows the effectiveness of agentic reinforcement training. The {Qwen3-30a3} model shows a remarkable improvement on {SWE-Bench Multilingual}, surging from a baseline of approximately 32\% to a peak of 42.0\%. This substantial 10-point absolute gain underscores the power of RL to enhance generalization across diverse, multilingual tasks. 


\paragraph{Applied to Qwen3-Max-Thinking}

To push the boundaries of agentic coding performance, we applied our methodology to our flagship {Qwen3-Max-Thinking} model. The resulting model achieved a performance of {75.3\%} on SWE-Bench Verified. This result validates the effectiveness of our large-scale data generation pipeline at a production level, showcasing its ability to elevate state-of-the-art models.
\section{Related Work}

\paragraph{Agentic and Synthetic Environment Generation}
A significant line of research focuses on generating software engineering tasks synthetically to create scalable training data, often without relying on authentic historical issues. For instance, {SWE-smith}~\citep{swe-smith} procedurally generates tasks by artificially injecting bugs, while {SWE-Flow}~\citep{swe-flow} leverages test documentation to synthesize novel problems. Other works generate complex bugs from scratch~\citep{sonwane2025bugpilot} or create interactive tasks from bug reports~\citep{jin2023inferfix}. The scope of synthesis extends to generating command-line tasks from natural language~\citep{lin2018nl2bash,gandhi2026endless}. Similarly, {RepoST}~\citep{xie2024codebenchgen} employs synthetic tests to generate training data. While these methods provide scalable training data, they do not fully capture the complexity and long-tail challenges of real-world software issues. In contrast, our work focuses exclusively on building executable environments for real-world software engineering problems derived directly from public repositories.

\paragraph{Real-World Software Environment Setup}

Much recent work focuses purely on the environment configuration capability itself, proposing scripted, agentic, or expert-driven methods to create executable environments, but without providing verifiers for specific software issues \citep{hu2025repo2run, Milliken2025a, Bouzenia2025, horton2019dockerizeme, jain2024r2e, vergopoulos2025automated, guo2026evoconfig, setupbench2025, eliseeva2025envbench, kuang2025process, fu2025multi}. Another line of work provides end-to-end verification for real software issues. The seminal {SWE-bench}~\citep{swe-bench} established this paradigm with over 2,400 real-world Python issues, each with a test script to verify a fix. This Python-centric approach was significantly scaled by {SWE-rebench}~\citep{swe-rebench}, which developed a fully automated pipeline to generate over 21,000 verifiable tasks. Other works like {SWE-Gym}~\citep{pan2024training}, {SWE-bench-Live}~\citep{zhang2025swe}, {SWE-Factory}~\citep{guo2025swe}, {daVinci-Dev}~\citep{zeng2026davinci}, {SWE-Bench++}~\citep{wang2025swe}, and {SWE-bench+}~\citep{aleithan2024swe} have also aimed to expand data generation, primarily within the Python ecosystem. A few recent efforts have created multi-language benchmarks, such as {Multi-SWE-bench}~\citep{multi-swe-bench} and {SWE-PolyBench}~\citep{rashid2025swe}; however, these have generally been limited in scale. Our work systematically overcomes these limitations by presenting a highly scalable and reliable pipeline that operates across arbitrary languages and repositories, successfully generating over million-scale executable environments with a validated, task-specific verifier.

\paragraph{Building Code Verifier}
Automated code verifiers are typically constructed by executing test suites to validate solutions. Traditional test generation methods, including probability-based approaches~\citep{4222570}, constraint-based~\citep{6693084}, and search-based~\citep{harman2010testing, pynguin}, often suffer from limited coverage and poor readability, and are typically restricted to regression or implicit oracles~\citep{oracle_survey}. Recent work leverages LLMs to generate unit tests for verification~\citep{a3test, ChatUniTest, ut_survey, ut_gpt, chen2024b4}, though the resulting tests can be unreliable due to confidently incorrect assertions. CodeRM~\citep{ma-etal-2025-dynamic} improves reward signal quality by scaling the number of generated tests and dynamically adapting test counts to problem difficulty. In this work, we leverage a building agent to automatically generate verifier for complext repository-level software engineering issues.

\section{Conclusion}

In this paper, we introduced SWE-Universe, a scalable framework designed to overcome the critical bottlenecks of low yield, inconsistent quality, and prohibitive cost in generating real-world software engineering environments. Our autonomous building agent, powered by a custom-trained model and equipped with iterative self-verification and in-loop hacking detection, successfully constructed a massive dataset of over 800,000 executable, multi-lingual tasks—the largest of its kind. We demonstrated the profound value of this resource through large-scale agentic training, showing that it provides a powerful signal for both supervised learning and reinforcement learning, ultimately enabling our {Qwen3-Max-Thinking} model to achieve a score of 75.3\% on SWE-Bench Verified.

\clearpage

\clearpage
\bibliography{biblio}
\bibliographystyle{colm2024_conference}

\end{document}